\documentclass[a4paper,fleqn,usenatbib]{mnras}

\usepackage{newtxtext,newtxmath}

\usepackage[T1]{fontenc}
\usepackage{ae,aecompl}

\usepackage{graphicx}	
\usepackage{amsmath}	
\usepackage{amssymb}	

\newcommand{\msun}{{M$_{\odot}$}}

\title[{\em SOFIA} observations of SN\,1987A in 2016]{{\em SOFIA} mid-infrared observations of Supernova\,1987A in 2016 --- forward shocks and possible dust re-formation in the post-shocked region?}

\author[M. Matsuura et al.]{
Mikako Matsuura,$^{1}$ \thanks{E-mail: matsuuram@cardiff.ac.uk (MM)}
James M. De Buizer,$^{2}$ 
Richard G. Arendt,$^{3, 4}$ 
Eli Dwek,$^{3}$ 
             \newauthor 
M.J. Barlow,$^{5}$ 
Antonia Bevan,$^{5}$ 
Phil Cigan,$^{1}$
Haley L. Gomez,$^{1}$
Jeonghee Rho,$^{2, 6}$ 
            \newauthor 
Roger Wesson,$^{5}$ 
Patrice Bouchet,$^{7,8}$ 
John Danziger,$^{9}$ 
Margaret Meixner,$^{10, 11}$ 
\\
$^{1}$ School of Physics and Astronomy, Cardiff University, Queen's Buildings, The Parade, Cardiff, CF24 3AA, UK \\
$^{2}$ Stratospheric Observatory For Infrared Astronomy NASA Ames Research Center, MS 232-12 Moffett Field, CA 94035, USA\\
$^{3}$ Observational Cosmology Lab, Code 665, NASA Goddard Space Flight Center, Greenbelt, MD 20771, USA \\
$^{4}$ University of Maryland-Baltimore County, Baltimore, MD 21250, USA \\
$^{5}$ Department of Physics and Astronomy, University College London, Gower Street, London WC1E 6BT, UK \\
$^{6}$ SETI Institute, 189 N. Bernardo Avenue, Mountain View, CA 94043, USA \\
$^{7}$ DRF/IRFU/DAp, CEA-Saclay, F-91191 Gif-sur-Yvette, France \\
$^{8}$ NRS/AIM, Université Paris Diderot, F-9119, Gif-sur-Yvette, France \\
$^{9}$ Osservatorio Astronomico di Trieste, Via Tiepolo 11, Trieste, Italy \\
$^{10}$ Space Telescope Science Institute, 3700 San Martin Drive, Baltimore, MD 21218, USA \\
$^{11}$ Department of Physics and Astronomy, The Johns Hopkins University, 366 Bloomberg Center, \\
3400 N. Charles Street, Baltimore, MD 21218, USA \\
}

\date{Accepted XXX. Received YYY; in original form ZZZ}

\pubyear{2018}

\begin{document}
\label{firstpage}
\pagerange{\pageref{firstpage}--\pageref{lastpage}}
\maketitle

\begin{abstract}
The equatorial ring of Supernova (SN) 1987A has been exposed to forward shocks from the SN blast wave, 
and  it has been suggested that these forward shocks have been causing on-going destruction of dust in the ring.
We obtained {\em SOFIA FORCAST} 11.1, 19.7 and 31.5\,$\mu$m photometry of SN\,1987A in 2016. 
Compared with {\em Spitzer} measurements 10\,years earlier, the 31.5\,$\mu$m flux has significantly increased. 
The excess at 31.5\,$\mu$m appears to be related to the Herschel 70\,$\mu$m excess, which was detected 5 years earlier.
The dust mass needed to account for the the 31.5--70\,$\mu$m excess is 3--7$\times10^{-4}$\,\msun,  
more than ten times larger than the ring dust mass ($\sim1\times10^{-5}$\,\msun) estimate from the data 10-years earlier.
We argue that dust grains are re-formed or grown in the post-shock regions in the ring 
after forward shocks have destroyed pre-existing dust grains in the ring and released refractory elements into gas.
In the post-shock region, atoms can stick to surviving dust grains, and the dust mass may have increased (grain growth), 
or dust grains might have condensed directly from the gas.
 An alternative possibility is that the outer part of the expanding ejecta dust might have been heated 
by X-ray emission from the circumstellar ring. 
The future development of this excess could reveal whether grains are reformed in the post-shocked region of the ring or  eject dust is heated by X-ray.
\end{abstract}



\section{Introduction}

It has been proposed that core-collapse supernovae (SNe) play a dual role in the production and destruction of the dust in the interstellar media (ISM) of galaxies,
and currently, these contradictory roles are  subjects of intense investigations.
It has been proposed that a large mass (0.1--1\,\msun) of dust can be formed in SN ejecta, using  newly synthesised  elements, 
thus,  SNe can be an important source of dust in the ISM of galaxies \citep{Nozawa:2003p29406, Morgan:2003p15480, Dwek:2011p29471}.
In parallel, theories have predicted that SN blast waves should destroy ISM dust grains by sputtering \citep[e.g.][]{Barlow:1978ik, schneider04, bocchio14}, with only larger grains surviving. 
\citet{Jones:1994p8385} found that SN shocks could destroy $\sim$95\% of ISM dust grains, resulting in a lifetime of dust grains in the ISM to be a few hundred million years. 
\citet{temim:2015bs} suggested much shorter dust lifetimes in the Magellanic Clouds (a few ten Myrs).
On the other hand, recent full hydrodynamical modelling  \citep{Silvia:2012br, Slavin:2015in} has found much longer ISM grain lifetimes ($>$1Gyr) against destruction by SN shocks. 
Although the lifetime of ISM dust is one of the keys in understanding dust evolution in galaxies, dust lifetimes are still uncertain \citep{Micelotta17}, owing to limited understanding of SN dust destruction processes.

Divided ideas about dust formation and destruction in SNe and SNRs are also found in observations.
SN1987A was the first SN in which dust formation was reported \citep{Danziger89}.
Since then dust formation has been reported in over ten SNe and a few SNRs \citep{Gall:2014dk, Matsuura17, 2018SSRv..214...63S}
with the inferred dust masses in young SNe typically of the order of $10^{-6}$ to $10^{-3}$\,\msun\, \citep[e.g.][]{Wooden:1993p29432, Bouchet:2004bs, Kotak:2009jv}.
Twenty three years after the explosion, a large mass ($\sim$0.5\,\msun) of cold ($\sim$22 K) ejecta dust was found by far-infrared observations in SN\,1987A \citep{Matsuura:2011ij}.
After this finding of a large dust mass, the evolution of the dust mass of SN\,1987A was re-visited, 
and now there is a debate as to whether such a large mass of dust was present in early days but the dust emission was optically thick and the inferred dust mass was underestimated \citep{Dwek:2015ds, Sluder:2016tn}. 
An alternative possibility is that the dust mass was indeed small at early times and increased  over time \citep{Wesson:2014gs, Bevan:2016fs}.

Dust destruction by SN remnants (SNRs) has been predicted by theories, but its measurement is challenging.
\citet{Lakicevic:2015iw} analysed dust in the SNRs in the Large Magellanic Cloud, and found that the ISM dust mass towards the SNRs tended to be lower than for the surrounding regions.
They claimed that SNRs may destroy more dust than they produce. 
However, their finding of a reduced dust mass towards the SNRs may be attributable to hot SNR emission overwelming cold ISM dust emission,
thus it may not be conclusive that the analysis shows dust destruction by SNRs \citep{Matsuura:2016cu, Micelotta17}.
Modelling {\em Spitzer}'s  \citep{Werner:2004jt}  emission of old Galactic SNRs (the Cygnus Loop and the Puppis A), \citet{Sankrit:2010ka} and \citet{Arendt:2010kw} 
estimated that about 35\,\% and 25\,\% of dust grains have been destroyed.
Meanwhile,  \citet{Lau:2015fg} suggested that dust had survived the reverse shock in the $\sim$10,000  year old Galactic SNR, Sgr A East. 
Following the detection of CO molecules in the reverse shock region of the Galactic SNR, Cassiopeia A \citep{Rho:2009gc, Wallstrom:2013he}, chemical models have predicted that CO molecules can re-form in the post-shock regions but that it would be difficult to re-form dust in this SNR \citep{Biscaro:2014kh}.

The explosion of SN 1987A was detected in the Large Magellanic Cloud, that lies only 50 kpc away. 
Due to its close distance, SN 1987A provides a unique opportunity to monitor at almost all wavelengths how the SNR has evolved over the past 30 years. 
{\em Hubble Space Telescope} (HST) optical images showed that the SNR is composed of ejecta,  an equatorial circumstellar ring and two fainter outer rings.
The ring is thought to consist of the material lost from the progenitor via a stellar wind when the star was in the red-supergiant phase 20,000--40,000 years ago \citep{Arnett:1989p29666, McCray:1993p29839}. 
While the ejecta are expanding at about 2000\,km\,s$^{-1}$ on average, the equatorial ring (hereafter the ring) expands much more slowly (about 10--100\,km\,s$^{-1}$). 
The HST monitoring program captured the ejecta expansion as its appearance changed from a single blob in the 1990s to a keyhole shape in the 2000s. 
Finally, the fastest part of the forward shock has passed the ring, with shock heated material just outside the ring in a 2014 image \citep{Fransson:2015gp}.

\begin{figure*}
\centering
\resizebox{\hsize}{!}{\includegraphics{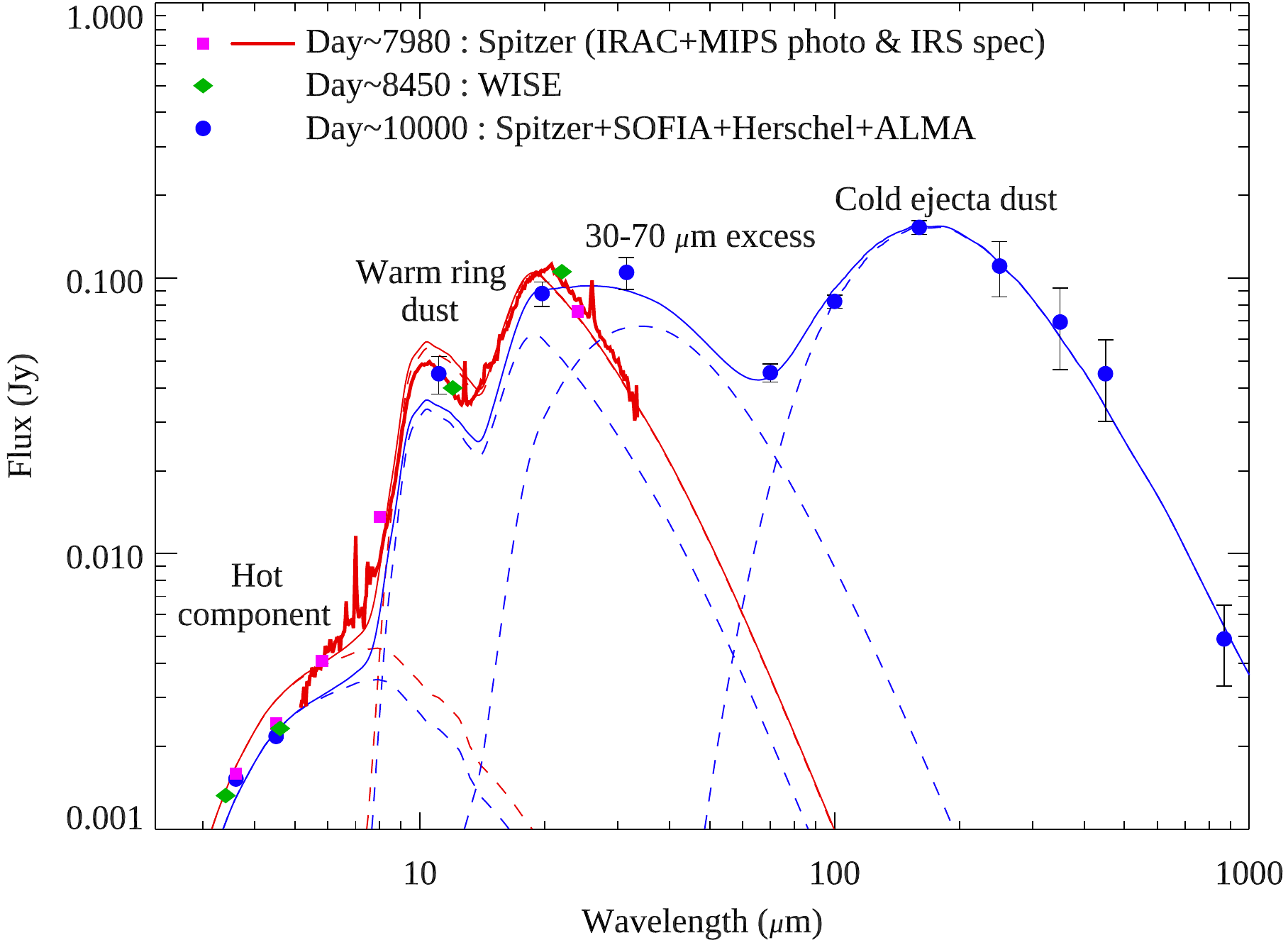}}
\caption{ The near- to far-infrared SED of SN\,1987A at three different epochs, with fitted dust models.
The pink squares and thick red line show the {\em Spitzer} IRAC and MIPS photometry data and IRS spectrum at day$\sim$7,980 \citep{Dwek:2010kv},
with the model fits as thin red lines (Model 1 in Table\,\ref{tab:dustmodel}).
{\em WISE} data at day 8,450 are plotted as green diamonds.
The blue circles show flux measurements around day 10,000, assembled from
{\em Spitzer} IRAC 3.6 and 4.5\,$\mu$m data at day 10,377 \citep{Arendt:2016ds},
{\em Herschel} measurements at day 9,090 \citep{Matsuura:2015kn},
ALMA measurements at 850 and 450\,$\mu$m at day 9294 and 9351 \citep{Zanardo:2014gu},
and our {\em SOFIA} observations at day 10,732.
One-$\sigma$ uncertainties are plotted.
The blue lines show four dust-component fits to the SED at day$\sim$10,000,
with  individual components plotted as dashed blue lines (Model 2 in Table\,\ref{tab:dustmodel}).
\label{fig-sed}}
\end{figure*}

For SN\,1987A, dust is found not only in the ejecta but also in the equatorial ring.
\citet{Bouchet:2006p2168} obtained spatially resolved images at 11.7 and 18.3\,$\mu$m, and identified silicate warm dust ($\sim$180 K) emission (Fig.\ref{fig-sed}) mainly arising from the ring.
Additionally, {\em Spitzer Space Telescope} observations found continuous emission between 3.6 and 4.5\,$\mu$m, and this component is attributed to collisionally heated dust in the ring \citep{Dwek:2010kv}, with a temperature of $\sim$525\,K \citep{Arendt:2016ds}.
In contrast, ALMA resolved images revealed that cold  \citep[$\sim$22 K; ][]{Matsuura:2015kn} dust emission clearly originates from the ejecta \citep{Indebetouw:2014bt, Zanardo:2014gu}.
SN 1987A has at least three discrete dust components in terms of  temperature (Fig.\ref{fig-sed}), and two in terms of location: the ring and the ejecta.

Over the last 12 years, {\em Spitzer} has monitored emission from the ring dust at 6-month intervals \citep{Dwek:2010kv, Arendt:2016ds}. 
After its launch in 2003, {\em Spitzer} detected increasing fluxes from 3.6\,$\mu$m to 24\,$\mu$m. After the liquid helium ran out in 2009, {\em Spitzer} continued to monitor only at 3.6\,$\mu$m and 4.5\,$\mu$m, and has found an increasing trend of hot dust components until day$\sim$9000,
when the fluxes started decreasing \citep{Arendt:2016ds}.
{\em Spitzer} observations have provided unique insights into the interaction between the SN blast wave and pre-existing material in the ring.

The infrared emission of the ring arises from $\sim$180\,K silicate dust grains, collisionally-heated by the SN blast wave \citep{Bouchet:2006p2168}. 
The same collisions are also capable of destroying the dust by thermal sputtering \citep{Dwek:2008p28793}.
Recently, \citet{Arendt:2016ds} reported that while the hot component has reduced its 3.6 and 4.5\,$\mu$m fluxes since $\sim$2012, the X-ray radiation, which is the heating source of dust, remains constant. 
They proposed that some dust grains in the ring have been destroyed.
Theory  \citep{Dwek:2010kv} predicts that these dust grains are expected to be destroyed by sputtering within $\sim$1\,yr. The existence of a hot component over a period of more than three years suggests that ambient circumstellar material is continuously being swept up by the shocks, acquiring more circumstellar dust.
The monitoring of the ring dust emission has detected  changes over a more than ten year timescale.

Here, we report SOFIA  photometry observations of SN\,1987A's dust at 11.1, 19.7 and 31.5\,$\mu$m obtained in June 2016,
after resumption of mid-IR monitoring.

\section{Observations}

SN\,1987A was observed with the NASA Stratospheric Observatory For Infrared Astronomy \citep[SOFIA; ][]{2012SPIE.8444E..10Y}
in the 2016 Cycle 4 observing cycle 
during a deployment to the southern hemisphere where the aircraft was temporarily based in Christchurch, New Zealand. 
Observations of SN1987A were taken on two separate flights (Flight 318 and 320). 
These flight numbers correspond to taking off on 11th July 2016 (day 10,731 since the explosion of SN\,1987A) and 13th July 2016 (day 10,733).
The observations were performed with the FORCAST imager and spectrometer \citep{Herter:2012hv} using 
the FOR$_{-}$F111, FOR$_{-}$F197, and FOR$_{-}$F315 filters. FORCAST has a short wavelength camera that is optimised to observe at wavelengths 
from 5 to 25\,$\mu$m, and a long wavelength camera optimised for observations from 25 to 40\,$\mu$m. 
All imaging observations were performed using the dual channel mode of FORCAST, which employs a dichroic to allow imaging in both cameras simultaneously. 
The 31.5\,$\mu$m filter remained in the long wavelength camera for all observations, while the short wavelength camera was configured to observe either 
in the 11.1\,$\mu$m or 19.7\,$\mu$m filter. Both cameras have 256$\times$256 pixels, which after distortion correction yield an effective field of view of 3.4'$\times$3.2' with a pixel scale of 0.768''\,pixel$^{-1}$. 
Observations were performed using the ``Nod-Match-Chop'' mode (a standard thermal infrared chop-nod background subtraction technique), and were configured to have 45'' East-West chop and nod throws.

Though co-added and calibrated pipelined data products were produced by the SOFIA Data Cycle System, 
the preliminary investigation of the data did not show detections at the signal-to-noise level expected. In particular, while there was a clear detection of SN\,1987A at 31\,$\mu$m 
from the first flight, there was no clear detection of it at that wavelength on the second flight, even though the observing time in this filter was comparable on both flights. 
The instrumental sensitivities in these filters are predominantly correlated to the water vapour overburden, with the 31.5\,$\mu$m filter being the most negatively affected by high water vapour of the three filters used.  
At the time of these observations, the observatory's water vapour monitor was non-functional, and therefore there is no valid information in the data headers that could be used 
to deduce the observing conditions during the observations. In order to investigate this further, the raw data products were downloaded from the SOFIA data archive and 
processed with a custom {\sc IDL} software package. These raw files contain data from each chop and nod position separately, allowing one to measure statistics related to the background emission, 
and by measuring these statistics in all files, one can deduce the atmospheric conditions by using the background behaviour with time as a proxy. 
The first flight experienced a brief episode of highly elevated background emission, which was likely due to temporary high precipitable water vapour conditions, and thus data during this episode were discarded. 
The second flight was almost completely plagued by highly variable backgrounds, again likely due to unusually high precipitable water vapour conditions. 
In addition, this second flight had some telescope pointing issues that could not be corrected in the data. 
The combination of these issues made it difficult to salvage any of the 11.1 and 31.5\,$\mu$m data from that flight. 
After discarding all data flagged for problems from both flights, the remaining data were co-added, 
yielding final effective on-source exposure times in the three filters of 5900\,s at 11.1\,$\mu$m, 3900\,s at 19.7\,$\mu$m, 
and 5900\,s at 31.5\,$\mu$m. The final signal-to-noise measurements on the detection of SN1987A at these three wavelengths are 6 at 11.1\,$\mu$m, 11 at 19.7\,$\mu$m, and 8 at 31.5\,$\mu$m.   

While chopping and nodding removes the vast majority of background sky and telescope emission, the large field of view of FORCAST and limited stability of in-flight observations leads to the presence of some low-frequency coherent background noise structures in the imaging data. To mitigate this, the co-added images were cropped to 78''$\times$78'', and by using a custom sky subtraction program, a low-power (between 2 and 5), two-dimensional, polynomial surface was fitted to the background with the central source being masked out. By subtracting these background surface fits from the co-added images, the final images were created at each of the three wavelengths.

The same flux calibrations applied to the pipeline-processed data were used to calibrate the final re-processed images. The SOFIA Data Cycle System pipeline calculates the flux calibration factors (i.e. Jy/ADU/sec) and errors based upon standard star observations across multiple flights and observing cycles, taking into account airmass corrections for telescope elevation and aircraft altitude, and these values are given in the pipeline-processed data headers. 
The flux calibration errors given in the headers are: 2.8\% at 11.1\,$\mu$m, 4.2\% at 19.7\,$\mu$m, and 7.0\% at 31.5\,$\mu$m. 
However, the dominant source of error in the flux density estimates in these particular data comes from the measurement errors due to the low S/N of SN\,1987A observations. 
The measurement errors for the aperture photometry are: 15.6\% at 11.1\,$\mu$m, 8.8\% at 19.7\,$\mu$m, and 11.9\% at 31.5\,$\mu$m. 
Therefore, the total flux calibration errors are these two values added in quadrature for each filter.
Applying these calibration factors and errors to the data produces the following measured flux densities and 1-sigma errors for SN1987A:  
45$\pm$7\,mJy at 11.1\,$\mu$m, 88$\pm$9\,mJy at 19.7\,$\mu$m, and 105$\pm$14\,mJy at 31.5\,$\mu$m (Table\,\ref{tab:fluxes}).

\begin{figure}
	\includegraphics[width=\columnwidth]{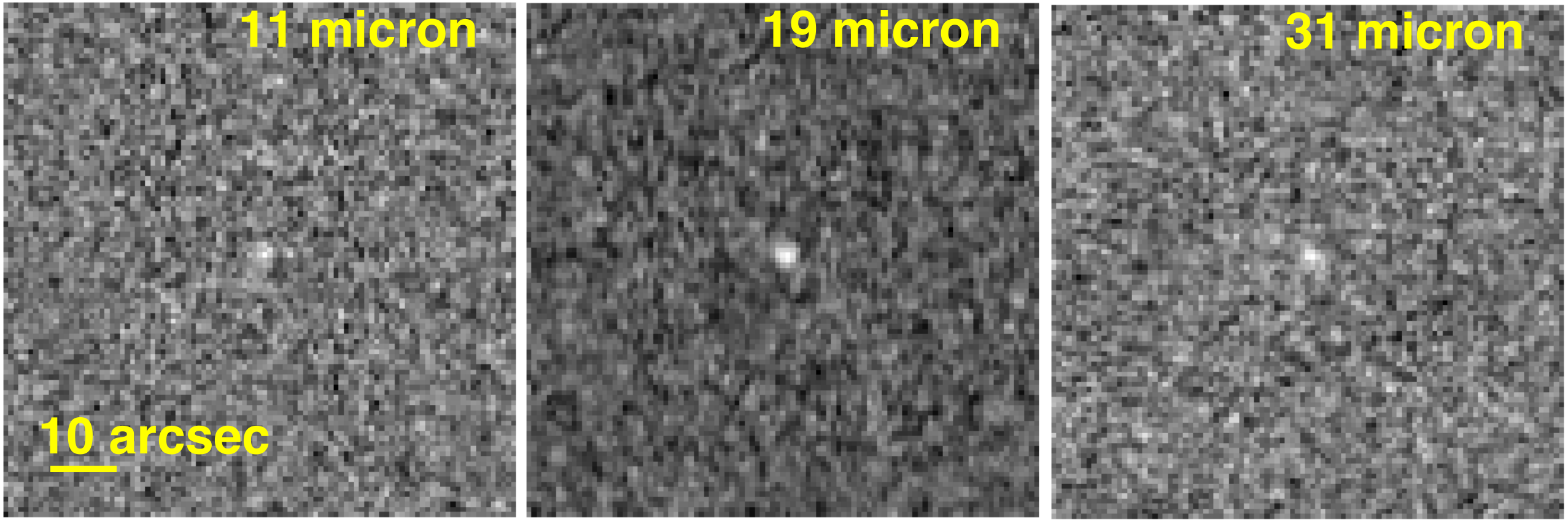}
    \caption{SOFIA images of SN 1987A, showing detections of an unresolved point source in all three filter bands.}
    \label{fig:images}
\end{figure}

Figure~\ref{fig:images} shows the SOFIA reduced images of SN\,1987A at three bands, with detections in all bands.
SN\,1987A is unresolved, seen as a point source.
This is expected because the full width of the half maximum (FWHM) of the point spread function at our shortest wavelength (11.7\,$\mu$m) is about 2.7\,arcsec (SOFIA Observing Handbook), which is larger than the size of the ring ($\sim$1.5\,arcsec in diameter).
The detection limit (S/N=4) of FORCAST in the 31.5\,$\mu$m band is estimated to be  105\,mJy for 5900\,sec exposure time in dual channel mode (SOFIA observer's Handbook for Cycle 4).
That is consistent with our detected flux - if the 31.5\,$\mu$m band flux had not increased since the Spitzer observations,
the source would be about 40\,mJy, and it would not have been detected at 31.5\,$\mu$m. Therefore,
 SN\,1987A has brightened at 31.5\,$\mu$m, allowing the source to be detected at this wavelength.

\begin{table}
	\centering
	\caption{{\em SOFIA} and {\em WISE} measured fluxes of SN\,1987A.}
	\label{tab:fluxes}
	\begin{tabular}{lrlr{@}{$\pm$}l} 
		\hline
		Filter & $\lambda_{\rm eff}$ & $\Delta \lambda$ & \multicolumn{2}{c}{Flux} \\
		         & $\mu$m & $\mu$m &  \multicolumn{2}{c}{mJy} \\
		\hline
                 \multicolumn{5}{l}{{\em SOFIA} day 10,732} \\
		FOR$_{-}$F111 & 11.1  & 0.95 & 45 & 7\\
		FOR$_{-}$F197 & 19.7 & 5.5 &  88 & 9\\
		FOR$_{-}$F315  & 31.5 & 5.7 & 105 & 14\\
		\hline
		\multicolumn{5}{l}{{\em WISE} mission between day 8330 and 8565} \\
		W1 & ~3.35 & 0.66 & 1.323 & 0.038 \\
		W2 & ~4.60 & 1.04 & 2.317 & 0.055 \\
		W3 & 11.56  & 5.51 & 39.95 & 0.68 \\
		W4 & 22.09  & 4.10 & 105.6 & 3.3 \\
		\hline
	\end{tabular}
\end{table}

\section{Spectral energy distributions}

\subsection{Historical spectral energy distributions}

Figure~\ref{fig-sed} shows the historical spectral  energy distributions (SEDs) of SN\,1987A.
The last {\em Spitzer} measurements before its helium ran out in 2009 are plotted in red in Fig.~\ref{fig-sed}.
The data include the {\em Spitzer} IRAC four photometric bands from 3.6 to 8.0\,$\mu$m at day 7974,
the MIPS 24\,$\mu$m flux at day 7983,
and IRS spectrum from 5 to 35\,$\mu$m at day 7954 \citep{Dwek:2010kv, Arendt:2016ds}.
\citeauthor{Dwek:2010kv} fitted the near- and mid-infrared SED  with two dust components, hot and warm (Fig.\,\ref{fig-sed}).

{\em Wide-Field Infrared Survey Explorer} (WISE) is an all-sky survey at 3.4, 4.6, 12 and 22\,$\mu$m \citep{Wright:2010in},
with the mission life from 14th December 2009 (day 8330) and 6th August 2010 (day 8565).
The magnitudes of SN 1987A were taken from the ALLWISE catalog Table\,\ref{tab:fluxes}, with the filter widths taken from \citet{Jarrett:2011jy}.
They were  converted to fluxes without colour corrections, yielding 
1.323$\pm$0.038, 2.317$\pm$0.055,   39.95$\pm$0.68,     105.6$\pm$3.3\,mJy
at 3.4, 4.6, 12 and 22\,$\mu$m, respectively.
The scanned dates of SN 1987A are not listed in the catalog, 
so we take the approximate date of the WISE scanned date as day 8450 since the explosion.


\subsection{The SED at day$\sim$10,000}

Figure~\ref{fig-sed} shows the {\em SOFIA} flux measurements of SN\,1987A at three bands at day$\sim$10,732.
In order to analyse these data points, we assembled the infrared and submillimeter flux measurements from dates close to the {\em SOFIA} observations.
The {\em Spitzer} warm mission measured the flux of SN\,1987A, using IRAC at 3.6 and 4.5\,$\mu$m, at day 10,377 \citep{Arendt:2016ds}.
The {\em Herschel Space Observatory} measured the flux of SN\,1987A at 70,  100, 160, 250 and 350\,$\mu$m at day 9,090 \citep{Matsuura:2015kn}.
The ALMA   fluxes of the ejecta were measured at 850 and 450\,$\mu$m at day 9,294 and 9,351 \citep{Zanardo:2014gu}. 
All these measurements are plotted in blue circles in Fig.\,\ref{fig-sed}.

Figure~\ref{fig-sed} demonstrates the presence of an excess at 31.5\,$\mu$m, which was not detected with {\em Spitzer} IRS observations ten years ago.
An excess at 70\,$\mu$m on top of the cold ejecta dust was reported by  \citet{Matsuura:2015kn}.
That could not be accounted  for by a potential contribution of 63\,$\mu$m [O\,{\small I}] line emission to the wide 70\,$\mu$m filter.
The excess found at  31.5\,$\mu$m appears to continue to the 70\,$\mu$m band, and we call this excess a `30--70\,$\mu$m excess'.

\citet{Arendt:2016ds} reported a decreasing trend for the hot dust components since day $\sim$8,500, after a long term increase since day 4,000.
This is also found in Fig.\,\ref{fig-sed} for  the {\em Spitzer} 4.5~$\mu$m and  {\em WISE} 4.6~$\mu$m fluxes.
The decreasing trend is also found in the {\em Spitzer} 3.6~$\mu$m flux.

In contrast to the hot component, the time variation of the warm component at 8--20\,$\mu$m is unclear.
The {\em SOFIA} 11.1\,$\mu$m flux at day 10,732 is consistent with the {\em Spitzer} IRS spectrum at day 7,954
and the {\em WISE} 12\,$\mu$m at day $\sim$8450, with the consideration that these bands are on the shoulder of silicate emission at 10\,$\mu$m.
The {\em Spitzer} MIPS 24\,$\mu$m flux and the {\em WISE}  22 $\mu$m flux are also consistent with the  {\em Spitzer} IRS spectra. 
Figure\,\ref{fig-sed} includes a 1\,$\sigma$ uncertainty in plotting the  {\em SOFIA} 19.7\,$\mu$m flux.
Although the {\em SOFIA} 19.7\,$\mu$m flux at day 10,732  have decreased more than 1\,$\sigma$ uncertainty since the {\em Spitzer} IRS spectrum were taken at day 7,954,
they are still consistent within 3\,$\sigma$ uncertainties.

\section{Analysis}

\subsection{Modified black body fitting to the SED of day $\sim$10,000}

\begin{table*}
	\begin{center}
	\caption{Dust model parameters with modified black bodies}
	\label{tab:dustmodel}
	\scriptsize
	\begin{tabular}{lc r{@}{$\pm$}l r{@}{$\pm$}l     r{@}{$\pm$}l r{@}{$\pm$}l  r{@}{$\pm$}l  r{@}{$\pm$}l r{@}{$\pm$}l   r{@}{$\pm$}l r{@}{$\pm$}l lllll} 
		\hline
                 & 
		& \multicolumn{4}{c}{Hot Component} 
		& \multicolumn{6}{c}{Warm Component}
		& \multicolumn{4}{c}{30--70\,$\mu$m Excess} 
		& \multicolumn{4}{c}{Cold Component} 
		& Figure\\ 
        & 
        & \multicolumn{2}{c}{$M_d$}   & \multicolumn{2}{c}{$T_d$}  
        & \multicolumn{2}{c}{$M_d$}   & \multicolumn{2}{c}{$T_d$} & \multicolumn{2}{c}{q$^{(1)}$}
        & \multicolumn{2}{c}{$M_d$}   & \multicolumn{2}{c}{$T_d$} 
        & \multicolumn{2}{c}{$M_d$}   & \multicolumn{2}{c}{$T_d$} \\
        & 
        &  \multicolumn{2}{c}{$\times10^{-8}$(\msun)} & \multicolumn{2}{c}{(K)}  
        &  \multicolumn{2}{c}{$\times10^{-5}$(\msun)} & \multicolumn{2}{c}{(K)}  & \multicolumn{2}{c}{} 
        &  \multicolumn{2}{c}{$\times10^{-4}$(\msun)} & \multicolumn{2}{c}{(K)}  
        &  \multicolumn{2}{c}{(\msun)} & \multicolumn{2}{c}{(K)}  \\
		\hline
Model 1 & Day$\sim$7,980  &    2.69&0.07 & 525.250&0.008    & 0.900&0.005 &  190.98&0.02 & \multicolumn{2}{c}{--} & \multicolumn{8}{c}{}& Fig.\,\ref{fig-sed}\\
Model 2 & Day$\sim$10,000 &   2.0&0.3     & 525&17    & 0.6&0.7      &  191&11           & \multicolumn{2}{c}{--} &   3.6&2.0 & 85&4 &  0.549&0.08 & 20.3&0.5 & Fig.\,\ref{fig-sed}\\
Model 3 & Day$\sim$10,000 & 2.0&0.3 &    525&17              & \multicolumn{2}{l}{~~3.2}  &\multicolumn{2}{l}{~~~~150} &  \multicolumn{2}{c}{--} &  \multicolumn{4}{c}{} & 0.549&0.08& 20.3&0.5 & Fig.\,\ref{fig:sed-3}  \\
Model 4 & Day$\sim$10,000 &  2.0&0.2      & 525&13    & 0.6&0.3      &  191&5           & \multicolumn{2}{c}{--} &   3.9&8.0 & 85&17 & 0.493&0.10 & 20.4 & 0.3 & Fig.\,\ref{fig:sed-4}\\
Model 5 & Day$\sim$7,980  & 2.69&0.07     & 525.250&0.008        &1.6$^{(2)}$ &0.1 & 195$^{(2)}$ &1 & 3.9$^{(2)}$ & 0.3 & \multicolumn{8}{c}{} & Fig.\,\ref{fig:sed-2}\\
Model 6 & Day$\sim$10,000  & 2.0&0.3 & 525&17 & 74$^{(2)}$&40 &  187$^{(2)}$&15 & 2.4$^{(2)}$ &1.8 & \multicolumn{3}{c}{}    && 0.549&0.08 & 20.3&0.3 &  Fig.\,\ref{fig:sed-2}\\

		\hline
	\end{tabular}\\
\normalsize
\end{center}
\begin{flushleft}
 $^{(1)}$Unless specified, $q$ is fixed to 3.5.
The models 2 and 4 have an additional component to explain the 30--70\,$\mu$m excess:
the model 2 uses  only three components (hot, warm and 30--70\,$\mu$m excess) to fit 3.6--70\,$\mu$m fluxes, with fixed parameters of the cold component,
while the model 4 consider all four components as independent parameters, fitting 3.6--870\,$\mu$m fluxes.
The model 3 is to find a solution of the hot component in order to fit 11.1--31.5\,$\mu$m SOFIA data within 3-$\sigma$, instead of 1-$\sigma$ uncertainty.
The model 5 and 6 use temperature dependence of grain size for warm dust, and the temperature quoted here is for the highest temperature (for the smallest grains).
$^{(2)}$ : temperature gradient of collisionally heated grains was considered.
\end{flushleft}
\end{table*}

In order to interpret the 30--70\,$\mu$m excess, we outline the known dust components to the SED fit at day 10,000.
In the optically thin case, the flux density $F_{\nu}$ at the frequency $\nu$  from a dust mass ($M_d$)
 is given as a modified blackbody as 
\begin{equation}
  F_{\nu} = M_{d} \frac{4 \kappa_\nu \pi B_{\nu}}{4 \pi D^2}, 
   \label{eq:dust}
\end{equation}
where $M_d$ is the dust mass,
$B_\nu(T_d)$ is the Planck function,
and $T_d$ is the dust temperature  \citep{Hildebrand:1983tm}.
$D$  is the distance to the LMC, adopted to be 50\,kpc. 
The dust mass absorption coefficient $\kappa_{\nu, a}$ is expressed as
$  \kappa_{\nu, a} = 3 Q_{\nu}/4 \rho  a$,
where $\rho$ is the mass density of the dust grains, $Q_\nu$ is the dust emissivity at the frequency
$\nu$, and $a$ is the grain size.
In the Rayleigh limit, i.e. the grain size $a$ is much smaller than the emitting wavelength ($\lambda$),
 $a \ll \lambda$, 
 $\kappa_{\nu}$ can be simply expressed by a power-law 
 $ \kappa \propto \lambda ^ {-\beta}$, 
for spherical grains, without grain size dependence.
Thus, the flux $F_{\nu}$  becomes independent of the grain size, at a given dust temperature $T_d$.

For hot and warm components, \citet{Dwek:2010kv} and \citet{Arendt:2016ds} already made fits to the {\em Spitzer} day$\sim$7,980 fluxes.
 Using, the {\sc IDL} $\chi^2$ minimisation procedure, {\sc amoeba} \citep{2002nrca.book.....P}, we searched for parameters that can fit these  photometric bands and spectra with two components.
Uncertainties were estimated by Monte-Caro Method \citep{2002nrca.book.....P}.
These two-components (hot and warm)  were simultaneously fitted.
The fitted results and uncertainties were cross-checked with those with an independent {\sc IDL} fitting code, {\sc mpfit} \citep{2009ASPC..411..251M}.
The adopted parameters of the hot and warm components are summarised in  Models 1 and 2 of Table\,\ref{tab:dustmodel}, and plotted in Fig.\ref{fig-sed}.
\citeauthor{Dwek:2010kv} tested  four different types of dust compositions for the hot component,
and in our analysis, amorphous carbon \citep{Rouleau:1991p29668} is used.
The difference in the fitted temperature of the hot component at day $\sim$7000 is mainly due to assumed grain size:  \citeauthor{Dwek:2010kv}
estimated the grain size distribution from X-ray flux, while we fixed the grain size at 0.1\,$\mu$m, focusing on infrared flux only.
The IRS spectra clearly showed silicate features, and fitting of the warm component used the optical constants from 
\citet{Draine:1984p25590} and \citet{Laor:1993p29314}.
The derived parameters of the warm component is marginally different from those by \citeauthor{Dwek:2010kv}, showing little impact of the parameter difference of the hot component on those of the warm component.

Cold ejecta dust was reported by \citet{Matsuura:2015kn}, and  their {\em Herschel}  flux measurements were about four and half years
before the {\em SOFIA} measurements.
\citet{Matsuura:2011ij} suggested that the heating source of the ejecta dust is most likely due to $^{44}$Ti decay.
The half-life of $^{44}$Ti is estimated to be 85 years \citep{Jerkstrand:2011fz},
and the heating from $^{44}$Ti decay would have declined only by 4\,\% over this four and half year time.
Therefore, it is most likely that the luminosity of the ejecta dust  has changed little since the {\em Herschel} measurements within their uncertainties.
In Fig\,\ref{fig-sed}, the fitting with amorphous carbon \citep{Zubko:1996p29442} is plotted. 
 This is a fitting to Herschel 100--350\,$\mu$m and ALMA 450 \&850\,$\mu$m fluxes \citep{Zanardo:2014gu}, 
and an independent fitting from the hot and warm components.
\citet{Matsuura:2015kn} tested fitting the cold ejecta dust with amorphous silicates \citep{Jager:2003gl}, 
but the difference in the predicted 70\,$\mu$m fluxes between these two dust models is  negligible  (only a 4\,\% difference).
Thus, adopting different dust compositions does not affect our conclusion of having an excess at 70\,$\mu$m.

After fitting the warm ring dust and cold ejecta dust, the excess departs from 30 to 70\,$\mu$m at a $>6\sigma$ level.

\begin{figure}
\resizebox{\hsize}{!}{\includegraphics{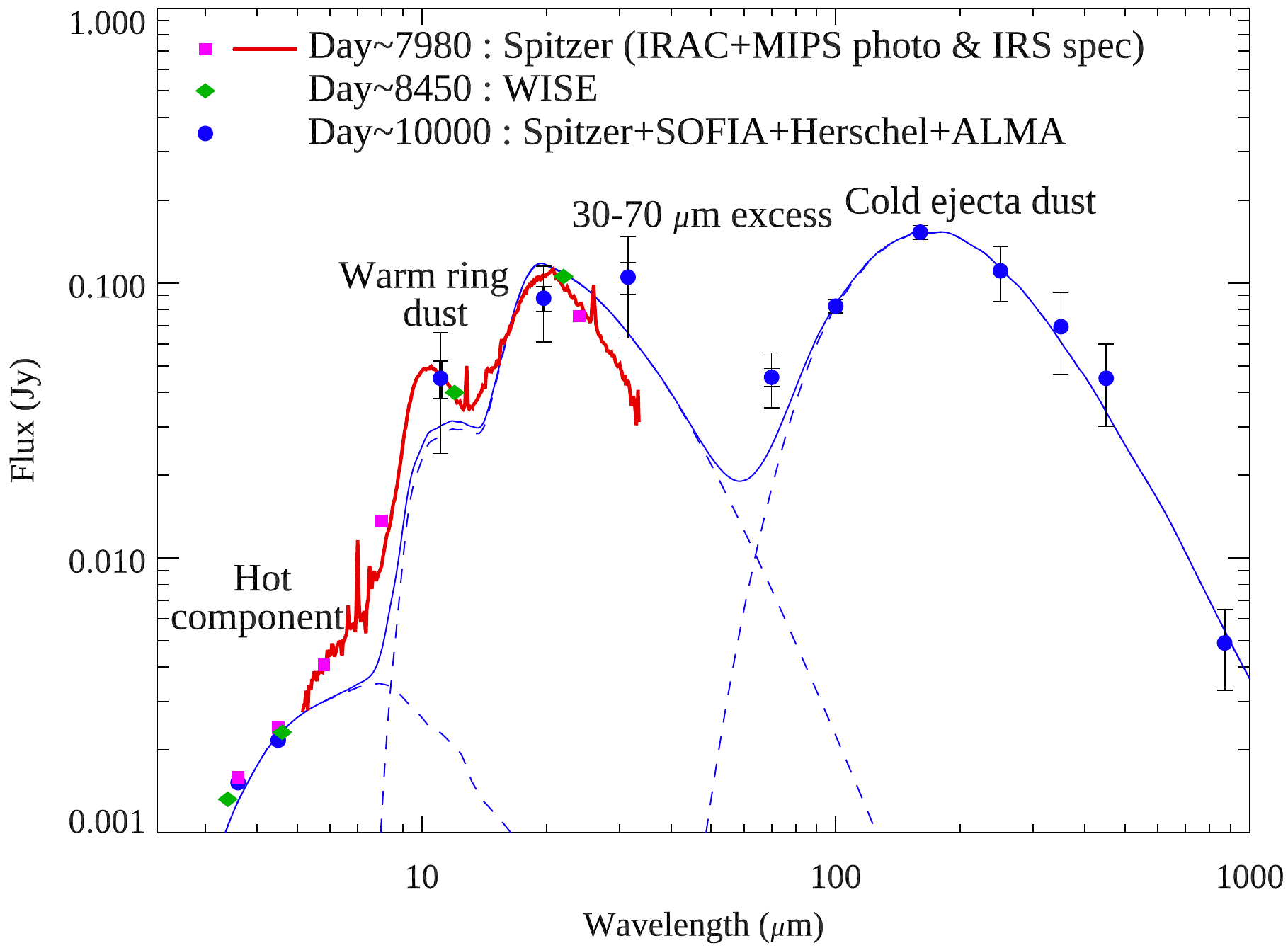}}
    \caption{Fitting the SEDs within  the 3-$\sigma$ uncertainties of the SOFIA fluxes  (Table\,\ref{tab:dustmodel}  model 3).
    Both 1-$\sigma$  and 3-$\sigma$ uncertainties are plotted for the SOFIA fluxes and Herschel 70\,$\mu$m,
    while the uncertainties for the Herschel and ALMA bands at $>$100\,$\mu$m remain 1-$\sigma$ only, the same as Fig.\,\ref{fig-sed}.
    The {\em Spitzer}'s uncertainties are smaller than the plotting symbols. 
    By increasing the dust mass and decreasing the dust temperature of the warm component, the {\em SOFIA} 11--31\,$\mu$m fluxes can be fitted
    within the 3-$\sigma$ uncertainties, but this fit is insufficient to reproduce the 70\,$\mu$m excess. }
    \label{fig:sed-3}
\end{figure}

In order to verify the presence of the excess, we further made 
 a single component fit to the SOFIA 11.1, 19.7 and 31.5\,$\mu$m fluxes,
considering their 3-$\sigma$ level uncertainties.
As is found in Fig.\,\ref{fig:sed-3},  a warm component with a dust temperature of  150\,K and a dust mass of  $3.2\times10^{-5}$\,\msun\,
 (Model 3 in Table\,\ref{tab:dustmodel})
can fit the SOFIA 11.1, 19.7 and 31.5\,$\mu$m fluxes within 3-$\sigma$ uncertainties.
In this fit, we vary only the parameters of the warm component, with those of hot and cold components fixed as of Model 2.
However, this model spectrum under-predicts the {\em Herschel} 70\,$\mu$m excess by a factor of 6; 
even with cold ejecta dust, this model still under-predicts the flux by a factor of 2. 
We further attempted to search for a fit to the {\em SOFIA} bands and the Herschel 70\,$\mu$m excess with the $\chi$-square minimisation procedure {\sc amoeba},
but no solution was found within the 3-$\sigma$ uncertainties.
A warm component with a modified blackbody fit is  insufficient to reproduce the 11--70\,$\mu$m fluxes.

\subsection{An additional `30--70-$\mu$m' component?}

In order to understand the nature of the 30--70 $\mu$m excess, we further model it with an additional modified blackbody component,  while we kept the parameters of hot, warm and cold as explained in the previous section.
The 30--70 $\mu$m excess 
requires the dust temperature to be 85\,K and the dust mass to be  $3.6\times10^{-4}$\,\msun\, (Table\,\ref{tab:dustmodel}  model 2; Fig.\ref{fig-sed})
assuming silicate dust  \citep{Draine:1984p25590}.
In this model, we fitted the `excess' on top of the warm and cold components. 
The temperature of 85\,K is between those of the warm and cold components.
The dust mass from the best fitted parameters is about a factor of 300 larger than that of the warm component.
Because the Planck Function is involved in equation \ref{eq:dust} and because lower dust temperature yields lower luminosity,  the $\sim$85\,K excess
would require a much higher dust mass than the $\sim$191\,K warm component, even though the uncertainty in the dust mass of the excess component is large.
%

\begin{figure}
\resizebox{\hsize}{!}{\includegraphics{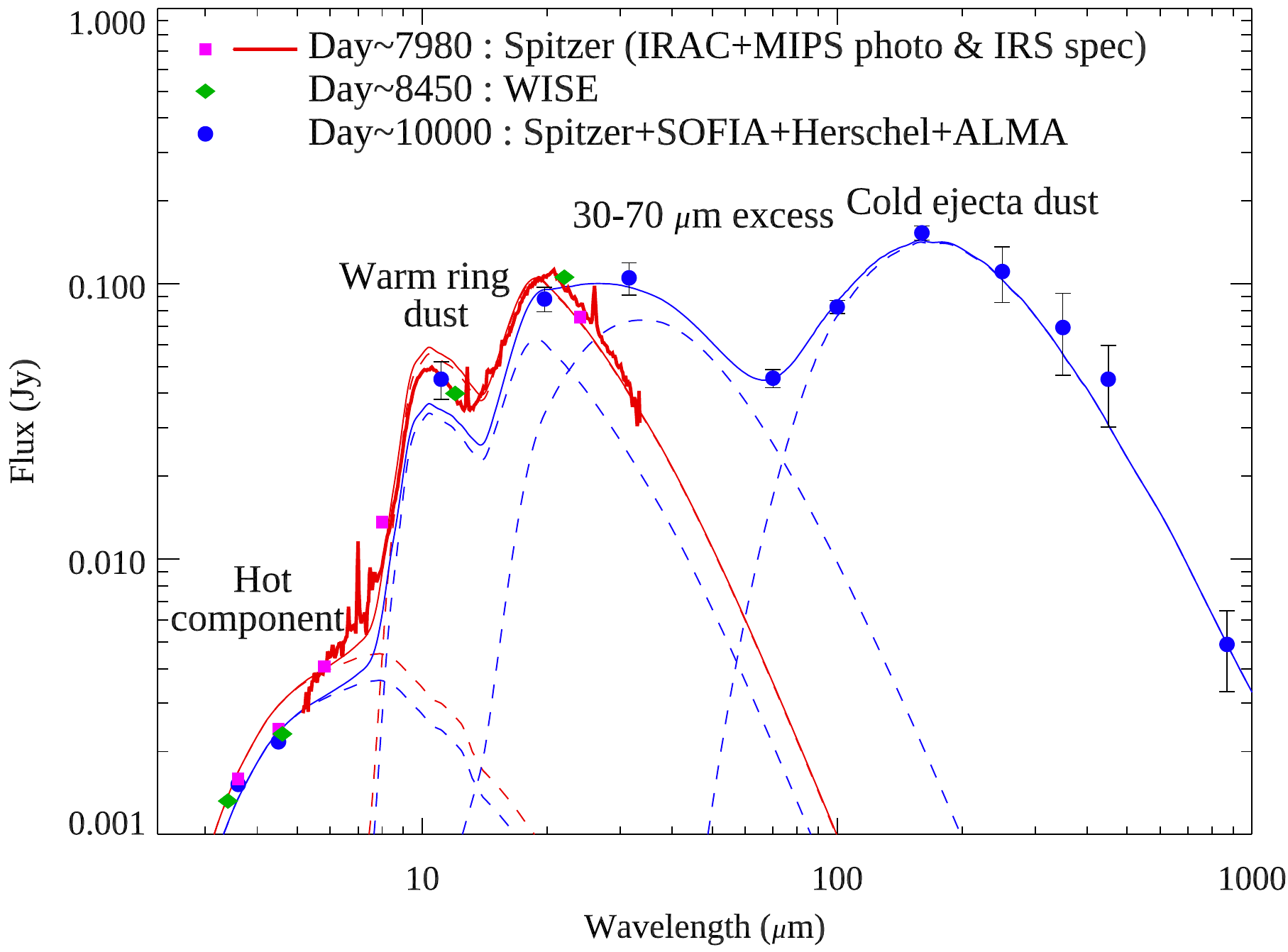}}
    \caption{SEDs and fitted results to day$\sim$10,000 with all four components as independent parameters (Table\,\ref{tab:dustmodel}  model 4)}
    \label{fig:sed-4}
\end{figure}

Finally, we fit all four components simultaneously as  free parameters, and the fitted results are summarised in Table\,\ref{tab:dustmodel}  model 4 and shown in Fig.\,\ref{fig:sed-4}. 
The best fitted parameters are consistent with the fitting of three components (hot, warm and cold) and an additional `30--70-$\mu$m' excess on top (Table\,\ref{tab:dustmodel}  model 2),
however, the fitting doesn't converge well, resulting in large uncertainties in the dust mass on the the excess component.

\subsection{Alternative possibility ---  temperature gradient and large grains in warm component?}

\subsubsection{Temperature gradient}

While Eq.\,\ref{eq:dust} assumes that all dust grains have the same dust temperature, this assumption might not be always the best, and may potentially be an oversimplification.
We consider an alternative possibility that dust grains in the   warm component have a temperature gradient as a function of grain size,
resulting in a wider spread in the emitting wavelengths.

The ring dust can be  continuously heated by X-ray radiation, emitted by the interaction between the fast SN blast wave
with the pre-existing equatorial ring \citep{Bouchet:2006p2168, Dwek:2010kv, Frank:2016ka}.
The fastest part of the ejecta gas expanded with a speed of over 3,000 km\,s$^{-1}$ \citep{McCray:1993p29839, Larsson:2016bj},
and has caught up with the slower \citep[100 km\,s$^{-1}$; ][]{Groningsson:2008jb} expanding ring.
In the shocked ring, there are two possible grain heating mechanisms, 
collisional heatings by shocked gas and radiative heating in the radiative shocks \citep{Bouchet:2006p2168}.
\citet{Dwek:2008p28793} mainly considered collisional heating
and noted that  the heating rate depends on whether electrons from the X-ray plasma stopped  inside the grains or `penetrate',
and that the grain temperature depends on grain size only when the electrons are trapped inside the grains.
Approximately, the temperature of a dust grain of a radius $a$ is a function of $a^{-\gamma}$, where $\gamma=1/(4+\beta)$.
Although this approximation is not accurate for the near-infrared,  the overall shape of  the SED at mid-infrared and far-infrared wavelengths does not have a significant impact due to this approximation.

First, we attempted to fit the SED from 10--70\,$\mu$m at day$\sim$10,000,  by including the grain size dependence of the dust temperature  in the warm component. The minimum and maximum grain sizes of 0.0003 and  1.0\,$\mu$m were adopted \citep{Weingartner:2001p3411}.
Such a model can fit the SED from 10--30\,$\mu$m within the 3-$\sigma$ flux uncertainties.
However, the model spectrum under-predicts the flux at 70\,$\mu$m, almost identically to the model spectra shown in Fig.\ref{fig:sed-3}.

\subsubsection{Temperature gradient with non-standard grain size distribution}

\begin{figure}
\resizebox{\hsize}{!}{\includegraphics{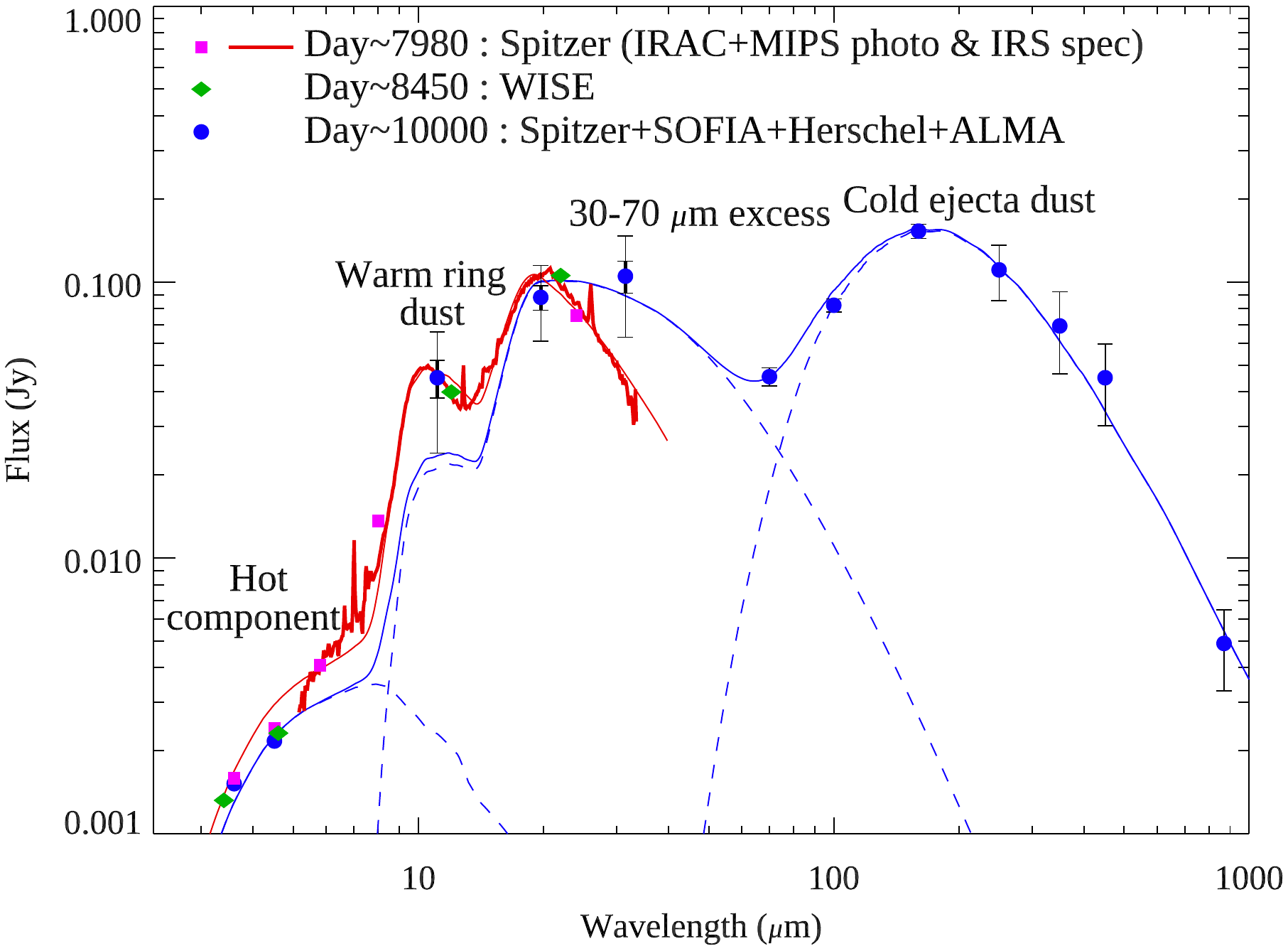}}
    \caption{Fitting the SEDs with models involving grain size distributions  ($q$) and the grain size dependent temperature for the warm component,  with parameters summarised 
    in Table\,\ref{tab:dustmodel}  models 5 and 6.
     The models for the hot and cold components remains the modified blackbody.}
    \label{fig:sed-2}
\end{figure}

 Although we added a simple function for the temperature dependence on grain size, 
using only  the power law index of $q$ of  3.5 for the Galactic ISM   \citep{Mathis:1977hp}, the fitted result under-predicted the flux at 70\,$\mu$m.
As the next step, we parameterise the power law index $q$ of the grain size distribution,
because increasing the number of larger dust grains can further increase the fluxes at longer wavelength.
We started by fitting  the warm component at day$\sim$8,000 with a power law index $q$ of 3.5.
The minimum and maximum grain sizes of 0.0003 and 1.0\,$\mu$m were adopted \citep{Weingartner:2001p3411}.
The maximum dust temperature (i.e. the dust temperature of the smallest grain) of  195\,K for the warm component
 can fit the SED at day$\sim$8,000  (Model 5 in Table\,\ref{tab:dustmodel}).
 During this process, the hot component still kept as Model 1, as having only two photometry points is insufficient to introduce additional parameter of $q$.
The resultant spectra are shown as red lines in Figure\,\ref{fig:sed-2}.

Using the $\chi^2$ fitting function {\sc amoeba}, we found that 
the spectra using a power law index $q$=2.4 with a maximum dust temperature of 187\,K can fit the 30--70\,$\mu$m excess  for day$\sim$10,000
 (Fig.\,\ref{fig:sed-2}).
 The fit is  slightly larger than 1-$\sigma$ uncertainty at 30\,$\mu$m but within 3-$\sigma$.
 We fixed the minimum and maximum grain sizes to be 0.003 and 1\,$\mu$m, as we have only four photometric points,
allowing optimisations of only up to three parameters (dust mass, temperature and the power law index of the grain size distribution).
The inferred dust mass was $7.4\times10^{-4}$\,\msun\, (Table\,\ref{tab:dustmodel} model 6).
 Although there is a large uncertainty in the dust mass and the inferred dust mass decreases with presence of more large dust grains (smaller $q$ index), 
 it is another issue whether such large dust grains, i.e. nearly flat grain size distribution across grain sizes, are plausible or not.

\section{Discussion -- origin of the 30--70\,micron excess}

 We have found  emission at 30--70\,$\mu$m in excess  on top of previously modelled composed of hot, warm and cold components. 
 We note that we consider it very unlikely that strong line emission could be a contributor to the rising 30\,$\mu$m flux.
 {\em Spitzer} IRS spectra did show a weak [S {\small III}]  line at 33.5\,$\mu$m,
which falls within the {\em SOFIA} 31.5\,$\mu$m band.
Additionally, [S {\small III}] has a line at 18.7\,$\mu$m.
If the increase of the {\em SOFIA} 31.5\,$\mu$m flux is due to [S {\small III}] line,
then the {\em SOFIA} 19.7\,$\mu$m band flux should have increased as well. However, such an increased trend was not found at 19.7\,$\mu$m.
It is unlikely that the line emission is the source of the increasing 30\,$\mu$m flux.

 We find two possibilities for the excess emission at 30--70\,$\mu$m; Model 2 an extra component at 30--70\,$\mu$m on top of the previously known hot, warm and cold components,
 or  Model 6 having a warm component with more large dust grains than the standard ISM grain size distributions.
In both cases, the  best fitted parameters of the 30--70\,$\mu$m emitting source have a dust mass higher than that of the warm component at day$\sim$8,000.
 We discuss the possible interpretations of these components and associated locations.

 
 If the excess is explained with an extra component, the excess can be fitted with a modified black body of an approximate temperature of 85\,K (Models 2 and 4).
The inferred dust mass is  $3.6\times10^{-4}$\,\msun\, and more than 10 times larger than the mass of the warm ring dust.
Because lower temperature of the modified black body results in lower brightness
(Eq.\,\ref{eq:dust}),
much higher dust mass in the excess component (temperature of $\sim$85\,K) would be required than that of the warm component (191\,K warm).
We argue that this excess might have originated from re-formed dust grains or from grain growth in the ring in the post-shock region.
Dust reformation refers to dust grains formed from the gas phase, whereas dust growth refers to dust grains with increased mass due to accretion of atoms from the gas phase.
Dust growth can accompany coagulation of grains with other dust grains, increasing grain sizes.

The fastest part of the blast wave started its interaction with the ring in 1995 \citep{Sonneborn:1998p29919},
and since then the morphology of the ring has changed in time;
initially the shape of the ring was smooth, but eventually it broke up into clumps, and now the blast wave (forward shock) has passed the ring 
\citep{Fransson:2015gp}.
The pressure behind the forward shock created reverse shock propagating into the SN ejecta \citep{Chevalier:2016te}.
The reverse shock caused by this interaction have been detected since 2003 \citep{France:2010p29071}.
As the forward shock expands outwards, the reverse shock moves inwards, the material left in between the two shocks can cool down,
and this material might be the site of dust re-formation.

A similar process but at a much earlier time frame has been proposed for type IIn SNe  \citep[e.g.][]{Smith:2008dz}.
Type IIn SNe show narrow line emission after the explosion, and these lines indicate the presence of dense circumstellar material that had been expelled by the progenitor star before the explosion.
The optical line asymmetries and the infrared excess have suggested formation of dust in the dense shell of
type IIn SNe, approximately a few ten to a few hundred days after the explosion 
\citep{Smith:2008dz, Fox:2009cl, Gall:2014dk, Andrews:2016jc, Chugai:2018jg}.
Similar dust formation in the material between the reverse and forward shocks might happen in SN\,1987A, but on a much longer time scale than for type IIn SNe.

Following the detection of CO molecules in the reverse shocked region in the Galactic SNR, Cassiopeia A \citep{Rho:2009gc, Rho:2012br}, 
chemical models have been developed to explain the presence of the CO molecules from ejecta material in the post-shocked region \citep{Biscaro:2014kh}.
However, the same chemical model predicts that dust is not easily formed in the post-shock regions.
 \citet{Biscaro:2014kh} modelled a type IIb SN,  and found that the density in the post-shock regions is not sufficiently high enough for dust formation.
 Furthermore, another chemical modelling \citep{2018ApJ...859...66S} of interaction between the circumstellar and SN blast winds (forward shock) in type IIn SNe found that the temperature in the post-shocked region is too high for early dust formation, when dust emission has been reported as early as day 87 in SN\,2010jl.
 If the density and the temperature are the key, that would open up a question of the density and temperature in post-shocked region in the circumstellar envelope in SN\,1987A. Estimate of the time evolution of the temperature and the density of SN\,1987A ring and chemical model on these physical conditions would be helpful if it is feasible for the dust formation in the SN\,1987A ring.


An alternative possibility to explain the 30--70\,$\mu$m excess is due to large dust grains (Model 6).
Compared with the power law index of $a^{-3.5}$ for the standard ISM dust grain distributions, where $a$ is the grain size,
the excess of SN\,1987A might be explained with $a^{-2.4}$, i.e. much more heavily weighted to larger dust grains than the standard ISM grain distribution.

SNR models have predicted that the power law index $q$ might depart from 3.5 in shocked SNRs
\citep{Nozawa:2007kh, Bianchi:2007p22212, Hirashita:2011jr}.
The model of \citet{Nozawa:2007kh} showed that smaller dust grains are likely to be destroyed by shocks,
while larger dust grains can survive.
It is possible that the grain size distribution
in the shocked circumstellar ring in SN\,1987A might not follow a standard ISM power law index of 3.5.

However, the inferred dust mass ($7.4\times10^{-4}$\,\msun) at day$\sim$10,000 is much larger than the mass
($1\times10^{-5}$\,\msun) at day$\sim$8,000.
That cannot be explained only by dust destruction processes,
and requires dust reformation or grain growth.


The inferred dust mass to explain the 30--70\,$\mu$m region has a larger dust mass than the warm component at day$\sim$8,000.
The timing coincides with passage of the forward shock beyond the ring, so it might be associated with forward shock.

It is unclear whether the excess found by {\em SOFIA} is dust re-formation or dust growth, as we are unable to disentangle these two cases from existing data.
After the passage of forward shocks into the ring of SN\,1987A, existing red-supergiant dust in the circumstellar ring should have been destroyed.
That would release refractory elements into gas.
Dust grains can be condensed from gas, i.e., re-forming dust grains in the post-shock region.
The passage of the forward shock will destroy dust grains in the circumstellar ring,  but  dust grains are  not completely destroyed, particularly larger grains.
Surviving dust grains could offer seeds for atoms to stick onto,
allowing the dust grain mass to increase with time.

The timescale for the dust reformation might be an issue.
The cooling timescale in the shocked gas in the ring of SN\,1987A has been predicted to be 12--40 years \citep{Dwek:2010kv}.
Since the forward shocks have been interacting until recently  \citep{France:2010p29071}, the cooling time scale
needs to be much shorter than that.

 An alternate possibility is that the 30--70\,$\mu$m excess is associated with the SN ejecta. 
The decreasing density of heavy elements will allow X-rays to heat an increasingly larger mass of ejecta dust to higher temperatures.
These possibilities will be tested by the forthcoming {\em JWST} space mission.
MIRI on board the {\em JWST} has a wavelength coverage up to 28\,$\mu$m, and it has an angular resolution sufficient to resolve
the inner structure of the ring and ejecta.
Thus, MIRI should be able to pin down exactly the location of the 30--70\,$\mu$m excess, whether it is within the ejecta,
or whether it is in dense material between the forward and reverse shock within the ring clumps.

If the dust mass can increase in post-shock regions, the  roles of SNe on dust evolution of galaxies might be re-evaluated.
Our observations suggest that it might be much easier to form dust grains than previously thought in a SN environment.
SNe have been discussed concerning their roles on largely destroying ISM dust with forward shocks, 
and as a source of ISM dust because newly formed dust can be formed from newly syntheised elements in the ejecta,
although such dust has been suggested to be completely destroyed by reverse shocks \citep{Nozawa:2007kh, Bianchi:2007p22212}.
Having dust forming in post-shock regions could potentially cause a re-evaluation of the the overall dust input from SNe and SN remnants
into the interstellar medium.

\section{Conclusions}

We report {\em SOFIA} flux measurements of SN\,1987A at 11.1, 19.7 and 31.5\,$\mu$m  in 2016.
We found that the 31.5\,$\mu$m flux has increased since  {\em Spitzer} measurements ten years ealier.
Together with  the excess found by  {\em Herschel} at 70\,$\mu$m, we consider the origin of  30--70\,$\mu$m continuum excess.
That excess can be fitted with dust component with a temperature of about 85\,K dust and with a dust mass of $3.5\times10^{-4}$\,\msun.
We suggest that the 30--70\,$\mu$m excess might be due to the dust re-formation in the circumstellar ring, after the passage of the forward shocks.
An alternative possibility is that part of the ejecta dust could be being heated to a much higher temperature than the rest of the ejecta dust.
If the 30--70\,$\mu$m excess is indeed due to dust re-formation,
that would suggest that dust formation or grain growth might  take place much  more easily and widely than previously thought.


\section*{Acknowledgements}

Based on observations made with the NASA/DLR {\em Stratospheric Observatory for Infrared Astronomy (SOFIA)}. {\em SOFIA} is jointly operated by the Universities Space Research Association, Inc. (USRA), under NASA contract NAS2-97001, and the Deutsches {\em SOFIA} Institut (DSI) under DLR contract 50 OK 0901 to the University of Stuttgart. 
Financial support for this work was provided by NASA through SOFIA 04-0016 issued by USRA.
This publication makes use of data products from the Wide-field Infrared Survey Explorer, which is a joint project of the University of California, Los Angeles, and the Jet Propulsion Laboratory/California Institute of Technology, funded by the National Aeronautics and Space Administration.
This work is based [in part] on archival data obtained with the {\em Spitzer Space Telescope}, which is operated by the Jet Propulsion Laboratory, California Institute of Technology under a contract with NASA. Support for this work was provided by an award issued by JPL/Caltech.
M.M.  acknowledges support from STFC Ernest Rutherford fellowship (ST/L003597/1),
M.J.B., A.B. and R.W. acknowledge support  from European Research Council (ERC) Advanced Grant SNDUST 694520, and 
HLG and P.C. acknowledge support from the European Research Council (ERC) in the form of Consolidator Grant {\sc CosmicDust} (ERC-2014-CoG-647939).



\bibliographystyle{mn2e}

\bibliography{sn1987a_sofia}




\bsp	
\label{lastpage}
\end{document}